
\input PHYZZX

\FIG\QUARKPOT{
The inter-quark potential $V(r)$ in the
dual Ginzburg-Landau theory.
The dashed curve denotes the Cornell potential.
}

\FIG\GBMASS{
The glueball masses $m_B(T)$ and $m_\chi (T)$ at
finite temperature.
A large glueball-mass reduction is found near $T_c$.
The phase transition occurs at the temperature satisfying
$m_\chi  \simeq T$, which is denoted by the dotted line.
}

\FIG\STRIN{
The string tension $k(T)$ at finite temperature $T$.
The lattice QCD result in the pure gauge
in Ref.[\gao] is shown by the dashed curve.
}

\FIG\TWOINST{
in the two-instanton system in the Polyakov gauge.
The two instantons with the same size are located at
$(\pm x_0,0,0,0)$ shown by small circles.
There appear two junctions and a loop in the QCD-monopole
trajectory.
}


\titlepage

\title{
Dual Ginzburg-Landau Theory for Nonperturbative QCD
\foot{
Invited Talk of the International Workshop on
``Color Confinement and Hadrons'', RCNP Osaka Japan,
March 1995, to appear in World Scientific Co.
}
}

\author{
H.~Suganuma, H.~Ichie, S.~Sasaki and H.~Toki
}

\address{
Research Center for Nuclear Physics (RCNP), Osaka University
\break
Mihogaoka 10-1, Ibaraki, Osaka 567, Japan
}

\abstract{
Nonperturbative QCD is studied with the dual Ginzburg-Landau
theory, where color confinement is realized through the
dual Higgs mechanism by QCD-monopole condensation.
We obtain a general analytic formula for the string tension.
A compact formula is derived for the screened inter-quark potential
in the presence of light dynamical quarks.
The QCD phase transition at finite temperature
is studied using the effective potential formalism.
The string tension and the QCD-monopole mass are largely reduced
near the critical temperature, $T_c$. The surface tension is
estimated from the effective potential at $T_c$.
We propose also a new scenario of the quark-gluon-plasma creation
through the color-electric flux-tube annihilation.
Finally, we discuss a close relation between instantons and
QCD-monopoles.
}

\chapter{Dual Higgs Mechanism for Color Confinement}

Color confinement and dynamical chiral-symmetry breaking
are quite outstanding features in the nonperturbative
\REF\kerson{
K.~Huang, {\it Quarks, Leptons and Gauge Fields},
(World Scientific, 1982).
}
\REF\rothe{
H.~J.~Rothe, {\it Lattice Gauge Theories}, (World Scientific,
1992).
}
QCD.$^{\kerson,\rothe}$
In particular, color confinement is extremely unique,
and is characterized by the formation of the color-electric flux tube
$^{\rothe}$ with the string tension about 1GeV/fm.
To understand the confinement mechanism,
much attention has been paid for
the analogy between the superconductor and the QCD
\REF\nambuA{
Y.~Nambu, {\it Phys.~Rev.}~{\bf D10} (1974) 4262.
\nextline
G.~'t~Hooft, in {\it High Energy Physics},
(Editorice Compositori, Bologna, 1975).
\nextline
S.~Mandelstam, {\it Phys.~Rep.}~{\bf C23} (1976) 245.
}
vacuum.$^{\nambuA}$
\REF\lifshitz{
E.~M.~Lifshitz and L.~P.~Pitaevsii,
{\it Statistical Physics},
(Pergamon press, 1981).
}
Similar to the superconductivity,$^{\lifshitz}$
the color-electric flux seems to be
excluded in the QCD vacuum, which leads the formation of the
squeezed color-flux tube between color sources.

In this analogy, color confinement is brought by the dual
Meissner effect originated from color-magnetic monopole condensation,
which corresponds to Cooper-pair condensation in the superconductivity.
As for the appearance of color-magnetic monopoles in QCD,
\REF\thooftB{
G.~'t~Hooft, {\it Nucl.~Phys.}~{\bf B190} (1981) 455.
}
't~Hooft$^{\thooftB}$
proposed an interesting idea of the abelian gauge
fixing, which is defined by the diagonalization of a
gauge-dependent variable.
In this gauge, QCD is reduced into
an abelian gauge theory with QCD-monopoles,
which appear from the hedgehog-like configuration corresponding to
the nontrivial homotopy class on the nonabelian manifold,
$\pi _2({\rm SU}(N_c)/{\rm U}(1)^{N_c-1})=Z_\infty ^{N_c-1}$.

We briefly compare the dual Higgs mechanism in the nonperturbative
QCD vacuum with the ordinary Higgs mechanism.
Like the Cooper pair in the superconductivity
or the Higgs field in the standard theory,
the charged-matter field to be condensed
is the essential degrees of freedom for the Higgs mechanism.
On the other hand, there is only the gauge field in the
pure gauge QCD, and hence it seems difficult to find any
similarity with the Higgs mechanism.
In the abelian gauge, however, only the diagonal gluon remains
as the gauge field, and the off-diagonal gluon behaves
as the charged-matter field, which leads QCD-monopoles
as the relevant degrees of freedom for color confinement.
Condensation of QCD-monopoles leads to mass generation
of the dual gauge field through the dual Higgs
\REF\suganumaA{
H.~Suganuma, S.~Sasaki and H.~Toki,
{\it Nucl.~Phys.}~{\bf B435} (1995) 207.
}
\REF\suzuki{
T.~Suzuki, {\it Prog.~Theor.~Phys.}~{\bf 80} (1988) 929 ;
{\bf 81} (1989) 752.
\nextline
S.~Maedan and T.~Suzuki, {\it Prog.~Theor.~Phys.}~{\bf 81} (1989) 229.
}
mechanism,$^{\suganumaA,\suzuki}$
which is the dual version of the Higgs mechanism.
Thus, the QCD vacuum can be regarded
as the dual superconductor after the abelian gauge fixing.

In this framework, the nonperturbative QCD is mainly
described by the abelian gauge theory with QCD-monopoles,
which is called as the abelian dominance.
Many recent studies based on the lattice gauge theory
have supported QCD-monopole condensation
and the abelian dominance in the maximal abelian
\REF\kronfeld{
A.~S.~Kronfeld, G.~Schierholz and U.~-J.~Wiese,
{\it Nucl.~Phys.}~{\bf B293} (1987) 461.
\nextline
T.~Suzuki and I.~Yotsuyanagi,
{\it Phys.~Rev.}~{\bf D42} (1990) 4257.
\nextline
S.~Hioki, S.~Kitahara, S.~Kiura, Y.~Matsubara,
O.~Miyamura, S.~Ohno and T.~Suzuki,
{\it Phys.~Lett.}~{\bf B272} (1991) 326.
\nextline
H.~Shiba and T.~Suzuki,
{\it Nucl.~Phys.}~{\bf B} (Proc.~Suppl.) {\bf 34} (1994) 182.
\nextline
A. Di Giacomo, {\it this Proceedings}.
}
gauge.$^{\kronfeld}$
The dual Higgs scheme predicts the existence of the dual gauge field
and the QCD-monopole
as the relevant degrees of freedom related to
color confinement.$^{\nambuA,\suganumaA,\suzuki}$
It can be proved that
these particles are color-singlet, so that
they can be observed as physical states.
The dual gauge field and the QCD-monopole appear
as a massive axial-vector glueball and a massive
scalar
\REF\suganumaB{
H.~Suganuma, S.~Sasaki and H.~Toki,
in {\it Quark Confinement and Hadron Spectrum},
Como, Italy, (World Scientific, 1995) p.238.
\nextline
S.~Sasaki, H.~Suganuma and H.~Toki, {\it ibid} p.241.
\nextline
H.~Toki, H.~Suganuma and S.~Sasaki,
{\it Nucl.~Phys.}~{\bf A577} (1994) 353c.
}
glueball,$^{\suganumaA,\suzuki,\suganumaB}$
corresponding to the weak vector boson
and the Higgs particle in the electro-weak standard theory.

\chapter{Dual Ginzburg-Landau Theory and Inter-Quark Potential}

We study the nonperturbative QCD using the dual Ginzburg-Landau (DGL)
\REF\maedan{
S.~Maedan, Y.~Matsubara and T.~Suzuki,
{\it Prog.~Theor.~Phys.}~{\bf 84} (1990) 130.
}
theory,$^{\suganumaA,\suzuki,\maedan}$
which is an infrared effective theory of QCD based on
the dual Higgs mechanism by QCD-monopole
condensation.$^{\suganumaA,\suganumaB}$
The DGL Lagrangian is described by
the diagonal gluon $\vec A^\mu  \equiv (A^\mu _3,A^\mu _8)$,
the dual gauge field $\vec B^\mu  \equiv (B^\mu _3,B^\mu _8)$
and the QCD-monopole field $\chi _\alpha
(\alpha =1,2,3)$,$^{\suganumaA,\suzuki}$
$$
\eqalign{
&{\cal L}_{\rm DGL}=
-{1 \over n^2} [n\cdot (\partial \wedge \vec A)]^\nu
[n\cdot ^*(\partial \wedge \vec B)]_\nu
-{1 \over 2n^2} (
[n\cdot (\partial \wedge \vec A)]^2 +
[n\cdot (\partial \wedge \vec B)]^2 )
\cr
&+\bar q (i\not \partial- e \vec H \cdot \not \vec A -m)q
+\sum_{\alpha =1}^3[|(i\partial_\mu -g \vec \epsilon _\alpha
\cdot \vec B_\mu )\chi _\alpha |^2
-\lambda (|\chi _\alpha |^2-v^2)^2]
}
\eqn\DGLlag
$$
in the Zwanziger
\REF\zwanziger{
D.~Zwanziger, {\it Phys.~Rev.}~{\bf D3} (1971) 880.
}
form,$^{\zwanziger}$ where the duality
of the gauge theory becomes manifest.
Here, $n_\mu$ corresponds to the direction of the Dirac string,
$e$ is the gauge coupling constant,
$g$ is the unit magnetic charge obeying the Dirac condition
$eg=4\pi $, and $\vec \epsilon_\alpha $ denotes the relative
magnetic charge of the QCD-monopole field
$\chi _\alpha $.$^{\suganumaA,\suzuki}$
The magnetic charge $g \vec \epsilon _\alpha $ is
pseudoscalar because of the extended Maxwell equation,
$\nabla  \cdot {\bf H}=\rho _m$.
Hence, the dual gauge field $\vec B_\mu $ is axial-vector.
In the absence of matter fields, one finds an exact dual relation
between $\vec A_\mu $ and $\vec B_\mu $ in the field equation,
$\partial \wedge \vec B = ^* (\partial \wedge \vec A)$.

In the DGL theory, the self-interaction of the QCD-monopole field
$\chi _\alpha $ is introduced to realize QCD-monopole condensation.
When QCD-monopoles are condensed,
the dual Higgs mechanism occurs, and
the dual gauge field $\vec B_\mu $ becomes massive, $m_B= \sqrt 3 gv$.
The color-electric field is then excluded in the QCD vacuum
through the dual Meissner effect, and is
squeezed between color sources to form the hadron flux tube.
The QCD-monopole also becomes massive as $m_\chi  = 2\sqrt\lambda v$.

As for the symmetry of the DGL theory,
there is the dual gauge symmetry [U(1)$_3$$\times $U(1)$_8$]$_m$
corresponding to the local phase invariance of the QCD-monopole
field $\chi _\alpha $ $^{\suganumaA,\suzuki}$
as well as the residual gauge symmetry [U(1)$_3$$\times $U(1)$_8$]$_e$
embedded in SU(3)$_c$.
The dual gauge symmetry leads to
the conservation of the color-magnetic flux.
In the QCD-monopole condensed vacuum,
the dual gauge symmetry [U(1)$_3$$\times $U(1)$_8$]$_m$ is
spontaneously broken,
and therefore the color-magnetic flux is not conserved.
On the other hand, the residual gauge symmetry
[U(1)$_3$$\times $U(1)$_8$]$_e$
is never broken in this process.$^{\suganumaA}$

We investigate the inter-quark potential
in the quenched level using the DGL theory.$^{\suganumaA}$
By integrating over $A_\mu$ and $ B_\mu$ in the partition
functional, the current-current correlation
$^{\suganumaA,\suzuki}$ is obtained as
$
{\cal L}_j=-{1 \over 2}\vec j_\mu D^{\mu \nu }\vec j_\nu
$
with the nonperturbative gluon propagator,
$$
D_{\mu \nu }=-{1 \over p^2}\left\{ g_{\mu \nu }+(\alpha _e-1)
{p_\mu p_\nu \over p^2} \right\}
-{1 \over p^2}
{m_B^2 \over p^2-m_B^2}
{1 \over (n\cdot p)^2}
\epsilon ^\lambda  \ _{\mu \alpha \beta }
\epsilon _{\lambda \nu \gamma \delta }n^\alpha n^\gamma
p^\beta p^\delta
\eqn\GLa
$$
in the Lorentz gauge.
Putting a static quark with color charge $^{\kerson,\suganumaA}$
$\vec Q$ at ${\bf x}={\bf r}$
and a static antiquark with color charge
$-\vec Q$ at ${\bf x}={\bf 0}$, the quark current is written as
$
\vec j_\mu (x)=\vec Qg_{\mu 0}\{\delta ^3({\bf x}-{\bf r})
-\delta ^3({\bf x}-{\bf 0})\}.
$
Because of the axial symmetry of the system
and the energy minimum condition.$^{\suganumaA}$,
one should take ${\bf n} // {\bf r}$,
which is also used in a similar context of the dual string
theory.$^{\nambuA}$
Then, one obtains the inter-quark potential
including the Yukawa and the linear parts,$^{\suganumaA,\suzuki}$
$
V(r)=-{\vec Q^2 \over 4\pi }\cdot {e^{-m_Br} \over r}+kr.
$

To derive the expression for the string tension $k$,
we consider an idealized long flux-tube system, where
the field variables can be described by the cylindrical coordinate
$r_T \equiv (x^2+y^2)^{1/2}$.
Like the Abrikosov vortex in the superconductivity,$^{\lifshitz}$,
one should consider the ``core" of the hadron flux tubes,
where the QCD-monopole condensate $\bar \chi $ almost vanishes.
The analysis of the field equation shows
$\bar \chi (r_T) \simeq m_\chi  v r_T$ inside the core ($r_T \lsim m_\chi
^{-1}$),
and $\bar \chi (r_T) \simeq v$ outside the core ($r_T \gsim m_\chi ^{-1}$).
Hence, we adopt the Lorentzian-type ansatz for the QCD-monopole field,
$$
\bar \chi ^2(r_T^2) \simeq v^2{m_\chi ^2r_T^2 \over 1+m_\chi ^2r_T^2}
\quad
\hbox{or}
\quad
\bar \chi ^2(p_T^2) \simeq v^2{m_\chi ^2 \over p_T^2+m_\chi ^2}
\eqn\LAx
$$
with $p_T \equiv (p_x^2+p_y^2)^{1/2} \simeq r_T^{-1}$.
In this case, we obtain an analytical expression
for the string tension,
$$
\eqalign{
k
=
{{\vec Q^2 \over 8\pi } m_B^2 m_\chi
\over \sqrt{m_\chi ^2-4m_B^2}}
\ln\left({m_\chi + \sqrt{m_\chi ^2-4m_B^2}
\over m_\chi - \sqrt{m_\chi ^2-4m_B^2}}\right)
=
{{\vec Q^2 \over 4\pi } m_B^2 m_\chi
\over \sqrt{4m_B^2-m_\chi ^2}}
{\rm arccos} {m_\chi  \over 2m_B}.
}
\eqn\STx
$$
For the type-II limit ($m_B \ll m_\chi $), one finds
$
k \simeq {\vec Q^2 m_B^2 \over 8\pi }\ln({m_\chi ^2 \over m_B^2})
$,$^{\suganumaA}$
corresponding to the well-known formula for
the energy per unit length of the
Abrikosov vortex in the type-II
superconductor.$^{\lifshitz}$

As for the parameter set,
the recent lattice QCD
\REF\ejiri{
Y.~Matsubara, S.~Ejiri and T.~Suzuki, {\it Nucl.~Phys.}
{\bf B}(Proc.~Suppl.) {\bf 34} (1994) 176.
}
studies$^{\ejiri}$ suggest $m_B \simeq m_\chi $,
which means the QCD vacuum corresponds to the dual-superconductor
of the type between type I and type II.
We show in Fig.1 the inter-quark potential
with the choice of $e=2.0$, $m_\chi =m_B=1.67{\rm GeV}$
corresponding to $\lambda= 29.4$ and $v= 0.154{\rm GeV}$,
which provide $k \simeq 0.9$GeV/fm for the
string tension and the radius of the hadron flux as
$R \simeq 0.12{\rm fm}$.
Here, we have included the correction coming from off-diagonal
gluons in the short-range part.

\chapter{
Infrared Screening Effect by Dynamical Light Quarks
}

We study the infrared screening effect
on the confinement potential due to light
quarks.$^{\suganumaA}$
For instance, a long hadron string can be cut through the
light $q$-$\bar q$ pair creation, and therefore
the inter-quark potential is saturated in the infrared region.
Such a tendency is observed in the lattice QCD with dynamical
\REF\wien{
W.~B\"urger, M.~Faber, H.~Markum and M.~M\"uller,
{\it Phys.~Rev.}~{\bf D47} (1993) 3034.
}
quarks.$^{\rothe,\wien}$

We estimate the $q$-$\bar q$ pair creation rate $w$
in the color-electric field inside the hadron flux tube,
which is formed between valence quarks.
Using the Schwinger
\REF\SGTA{
H.~Suganuma and T.~Tatsumi, {\it Phys.~Lett.} {\bf B269} (1991) 371;
\nextline
{\it Ann.~Phys.}~(N.Y.)~{\bf 208} (1991) 470;
{\it Prog.~Theor.~Phys.}~{\bf 90} (1993) 379.
}
formula$^{\suganumaA,\SGTA}$ in QCD,
we estimate the expectation value of the energy of the created
$q$-$\bar q$ pair as
$\langle E_{q\bar q} \rangle \simeq 0.85 {\rm GeV}$.
Since the energy $\langle E_{q\bar q} \rangle $ is supplied
by the missing length of the hadronic string,
the infrared screening length $R_{\rm sc}$ can be estimated from
$k R_{\rm sc} \sim \langle E_{q\bar q} \rangle $.
Thus, one obtains $R_{\rm sc} \sim 1{\rm fm}$,
which corresponds to a typical value of the hadron
size.$^{\suganumaA}$

The hadronic string becomes unstable against
the $q$-$\bar q$ pair creation
when the distance between the valence
quarks becomes larger than $R_{\rm sc}$.
This means the vanishing of the strong correlation between the
valence quarks in the infrared region, so that
the corresponding infrared cutoff,
$a \simeq R_{\rm sc}^{-1} \sim 200{\rm MeV}$,
should appear in the system.$^{\suganumaA}$
Taking account of the infrared screening effect,
we introduce the infrared cutoff
$a$ to the nonperturbative gluon propagator {\GLa} by replacing
${1 \over (n \cdot p)^2} \rightarrow
{1 \over (n \cdot p)^2 +a^2}$,$^{\suganumaA}$
because the non-local factor
$\langle x |{1 \over (n \cdot p)^2}|y \rangle$
provides the strong and long-range correlation as
the origin of the confinement potential.$^{\suganumaA}$
Here, this gluon propagator keeps the residual gauge symmetry.
Such a disappearance of the infrared double pole in
the gluon propagator in the DGL theory can be qualitatively
shown by considering the polarization diagram of quarks.

Using the above gluon propagator,
we obtain a compact formula$^{\suganumaA}$ for the quark
potential including the infrared screening effect by the
$q$-$\bar q$ pair creation,
$$
V_{\rm sc}(r)
=-{\vec Q^2 \over 4\pi }\cdot {e^{-m_Br}\over r}
+k \cdot {1-e^{-ar} \over a},
\eqn\POTc
$$
which exhibits the saturation
for the longer distance than $a^{-1} \simeq 1{\rm fm}$.
This formula for the screened quark potential has been used
not only for the lattice QCD results with light dynamical
quarks$^{\rothe}$,
but also for the phenomenological analysis of the hadron
\REF\chao{
Y.-B.~Ding, K.-T.~Chao and D.-H.~Qin, {\it Phys.~Rev.}~{\bf D51}
(1995) 5064.
}
decay.$^{\chao}$

\chapter{
QCD Phase Transition at Finite Temperature
}

We study the QCD vacuum at finite
\REF\kapusta{
J.~I.~Kapusta, {\it Finite-Temperature Field Theory},
(Cambridge University Press, 1988).
}
temperature$^{\kapusta}$ using the DGL
\REF\ichie{
H.~Suganuma, S.~Sasaki, H.~Toki, H.~Ichie,
{\it Prog.Theor.Phys.}~(Suppl.) in press.
}
\REF\ichieB{
H.~Ichie, H.~Suganuma and H.~Toki,
{\it Phys.~Rev.}~{\bf D} in press.
}
theory$^{\ichie,\ichieB}$ at the quenched level,
where the quark degrees of freedom are frozen.
In this case, the quark term can be dropped in the DGL Lagrangian,
and therefore one obtains a simple partition
functional$^{\ichie,\ichieB}$ after the integration
over the gauge field $A_\mu $,
$$
\eqalign{
  Z[J] &= \int {\cal D}{\chi_{\alpha}}{\cal D}{\vec{B}_{\mu}}
\exp{\left( i\int d^4x \{{\cal{L}}_{\rm DGL}^{\rm quench}
-J\sum_{\alpha=1}^3|\chi_\alpha|^2\} \right) },
\cr
 {\cal L}_{\rm DGL}^{\rm quench} &\equiv - {1 \over 4}
(\partial_{\mu}\vec{B}_{\nu}-\partial_{\nu}\vec{B}_{\mu})^2 +
 \sum_{\alpha=1}^3[|(i\partial_{\mu}-g\vec{\epsilon_{\alpha}}
\cdot\vec{B}_{\mu})
\chi_{\alpha}|^2 - \lambda(|\chi_{\alpha}|^2-v^2)^2].
}
\eqn\La
$$
Here, we have introduced the quadratic source
term.$^{\ichie,\ichieB}$
The thermodynamical potential is then obtained as
$$
\eqalign{
V_{\rm eff}(\bar \chi ;T) =   3 \lambda ( \bar \chi^2 - v^2 )^2
          &+ 3 {T \over \pi^2} \int_0^\infty  dk k^2 \ln{
           \left(  1 - e^{ - \sqrt{ k^2 + m_B^2}/T }
           \right)  } \cr
          &+ {3 \over 2} {T \over \pi^2}\int_0^\infty  dk k^2 \ln{
           \left(  1 - e^{ - \sqrt{ k^2 + m_{\chi}^2}/T }
           \right)  }.
}
\eqn\Vb
$$
The glueball masses $m_\chi $ and $m_B$ depend on
the QCD-monopole condensate $\bar \chi $,
$$
   m_\chi ^2(\bar \chi ) = 2\lambda (3 \bar \chi ^2-v^2) + J(\bar \chi )
   = 4\lambda \bar \chi ^2,
   \hbox{\quad} m_B^2(\bar \chi ) = 3 g^2 \bar \chi ^2,
\eqn\GBmas
$$
which are always non-negative for the whole region of $\bar \chi $.
Owing to the introduction of the quadratic source
term in Eq.{\La},  we can formulate the effective action for the
whole region of the order parameter without any difficulty
of the imaginary scalar-mass problem $^{\kapusta,\ichie,\ichieB}$
in the $\phi ^4$-type theory.

Like the Ginzburg-Landau theory in the superconductivity,$^{\lifshitz}$
one should consider the possibility of the $T$-dependence
on the parameters ($\lambda $,$v$) in the DGL theory.
In particular, the self-interaction of $\chi _\alpha $
is introduced phenomenologically,
and it should be reduced at high $T$ according to
the asymptotic freedom behavior of QCD.
Hence, we adopt a simple ansatz for the $T$-dependence
on $\lambda $,$^{\ichie,\ichieB}$
$
\lambda (T) \equiv \lambda  ( 1 - \alpha  T/T_c).
$
(We take $\lambda (T)=0$ for $T>T_c/\alpha $.)
We take $\alpha  \simeq 0.97$ to reproduce the thermodynamical
critical temperature $T_c \simeq 0.2$GeV,
which means a large reduction of the
self-interaction among QCD monopoles near $T_c$.
Our results to be shown below do not depend largely
on the value of $T_c$,
which may takes a larger value, e.g. $T_c \simeq 0.26$GeV,
suggested from the recent lattice
\REF\karschA{
J.~Fingberg, U.~Helle and F.~Karsch, {\it Nucl.~Phys.}
{\bf B392} (1993) 493.
}
QCD simulations$^{\karschA}$.

We find a first-order phase transition at $T_c$=0.2GeV.
The lower and upper critical temperatures are
$T_{\rm low}=0.163$GeV and $T_{\rm up}=0.205$GeV, respectively.
We show in Fig.2 the glueball masses
$m_B(T)$ and $m_\chi (T)$ at finite temperature.
A large reduction of the glueball mass is suggested near $T_c$.
In particular, the QCD-monopole mass $m_\chi (T)$ largely
drop down to $m_\chi  \sim T_c$ ($\simeq$ 0.2GeV) near $T_c$.
The deconfinement phase transition occurs at the
temperature satisfying $m_\chi  \simeq T$,
which seems quite natural because
only low-lying modes with $\omega _n \lsim T$ can contribute
significantly due to the thermodynamical factor
$
1 / (e^{\omega _n/T} \pm 1 )
$.
Similar glueball-mass reduction is also
suggested by the thermodynamical studies
based on the lattice
\REF\karsch{
J.~Engels, F.~Karsch, H.~Satz and I.~Montvay,
{\it Phys.~Lett.}~{\bf B102} (1981) 332.
}QCD data.$^{\karsch}$

We show in Fig.3 the string tension $k(T)$ at finite temperature,
calculated by using Eq.{\STx}.
The string tension $k(T)$ decreases rapidly with temperature,
and drops down to zero around $T_c = $ 0.2 GeV.
Hence, one expects a rapid change of
the masses and the sizes of the quarkonia
according to the large reduction of $k(T)$ near $T_c$.
Our result agrees with the lattice QCD data in the
\REF\gao{
M.~Gao, {\it Nucl.~Phys.}~{\bf B9} (Proc. Suppl.) (1989) 368.
}
pure gauge$^{\gao}$: $k(T) \simeq k(0)(1-T/T_c)^{0.42}$.

We estimate the surface tension $\sigma $
between the confinement and deconfinement phases
using the effective potential at $T_c$ in the DGL theory.
There are two minima at $\bar \chi  = 0$ and
$\bar \chi = \bar \chi _c$ in $V_{\rm eff}(\bar \chi ;T_c)$.
The mixed phase includes both the confinement phase
($\bar \chi =\bar \chi _c$) and the deconfinement phase ($\bar \chi =0$).
When the boundary surface in the mixed phase is taken on
the $xy$-plane ($z=0$),
the system depends only on the $z$-coordinate,
and the boundary condition is given as
$
\bar \chi (z=-\infty )=0, \quad \bar \chi (z=\infty )=\bar \chi _c.
$
The surface tension $\sigma $ in the DGL theory is estimated as
$$
\sigma  \simeq \int_{-\infty }^\infty  dz  \left\{
3 \left({d \bar \chi (z) \over dz} \right)^2
+V_{\rm eff}[\bar \chi (z);T_c] \right\}.
\eqn\SurT
$$
We approximate the figure of $V_{\rm eff}(\bar \chi ;T_c)$
($0 \le \bar \chi  \le \bar \chi _c$) as a sine curve,
\nextline
$
V_{\rm eff}(\bar \chi ;T_c)
\simeq  {h \over 2}
\{1-\cos (2\pi  \bar \chi  / \bar \chi _c)\},
$
where $h$ and $\bar \chi _c$ corresponds to ``height" and
``width" of the ``potential barrier" between the two stable
states at $T_c$.
The field equation of $\bar \chi (z)$
is then solved analytically like the sine-Gordon
\REF\rajaraman{
R.~Rajaraman, {\it Solitons and Instantons},
(North-Holland, 1982).
}
equation,$^{\rajaraman}$
$$
\bar\chi (z) \simeq {2\sqrt{6} \over 3} \tan^{-1} e^{z/\delta },
\quad
\delta  \equiv { \sqrt{3} \over \pi } \bar \chi _c/\sqrt{h},
\quad
\sigma  \simeq {4\sqrt{3} \over \pi } \sqrt{h} \bar \chi _c,
\eqn\SurTA
$$
where $\delta $ denotes the thickness of the boundary
between the two phases.
In terms of the effective potential,
the smallness of $\sigma $ corresponds to the weakness of the first-order
phase transition, because $\sigma $ takes smaller value for smaller
``height" $h$ or ``width" $\bar \chi_c$ of the
``potential barrier" in $V_{\rm eff}(\bar \chi ;T_c)$.

One finds $\bar \chi _c \simeq 0.75{\rm fm}^{-1}$ and
$h \simeq 0.33 {\rm fm}^{-4}$ from
$V_{\rm eff}(\bar \chi ;T_c)$.
Hence, the surface tension is estimated as
$\sigma^{1/3}  \simeq 196{\rm MeV}$,
and the thickness of the border between the two phases
is $\delta  \simeq 0.7{\rm fm}$.
Since the above estimation has been done in the quenched level,
the obtained results are to be compared with
the lattice QCD data in the quenched level, e.g.
\REF\iwasaki{
Y.~Iwasaki, K.~Kanaya, L.~K\"arkk\"ainen, K.~Rummukainen
and T.~Yoshie, {\it Phys.~Rev.}~{\bf D49} (1994) 3540.
}
$\sigma ^{1/3} \sim 80 {\rm MeV}$.$^{\iwasaki}$
Therefore, our estimation for $\sigma ^{1/3}$ seems
rather good in spite of the rough treatment.

\chapter{
Application to Quark-Gluon-Plasma Physics
}

We apply the DGL theory to the
quark-gluon-plasma (QGP) physics in ultrarelativistic
heavy-ion collisions.
In a standard picture of the QGP formation,
many color-electric flux tubes are
formed between heavy ions immediately after the
\REF\glendenning{
N.~K.~Glendenning and T.~Matsui,
{\it Phys.~Rev.}~{\bf D28} (1983) 2890;
\nextline
{\it Phys.~Lett.}~{\bf B141} (1984) 419.
}
collision.$^{\SGTA,\glendenning}$
In this pre-equilibrium stage,
there occurs $q$-$\bar q$ pair creation violently
inside tubes.$^{\SGTA,\glendenning}$, and
the energy of the color-electric field
turns into that of the stochastic kinetic motion
of quarks (and gluons).
The energy deposition and the thermalization thus occur.

We here examine the interaction between
the color-electric flux tubes in the DGL theory
to study the QGP formation
in terms of the flux-tube dynamics,
because many flux tubes would overlap
in the central region between heavy ions
just after collisions.
There are several kinds of flux tubes in the QCD system.
Each flux tube is characterized by the
color charge $^{\kerson,\suganumaA}$
$\vec Q$ at its end.

We study the interaction between two color-electric flux tubes
with the color-electric charge $\vec Q_1$ and $\vec Q_2$
at their ends.
The system is idealized as two sufficiently long flux tubes,
where the separation distance between them is denoted by $d$.
For $d \gg m_\chi ^{-1}$,
the interaction energy per unit length in this system
is estimated as$^{\ichie}$
$
E_{\rm int} \simeq  {\vec Q_1 \cdot \vec Q_2 \over 2\pi }
m_B^2 K_0(m_Bd)
$
using the similar calculation
for the Abrikosov vortex in the superconductor.$^{\lifshitz}$

There are two interesting cases
on the interaction between two color-electric flux tubes.

\item{\rm (a)}
For the same flux tubes with opposite flux
direction (e.g. $R$-$\bar R$ and $\bar R$-$R$),
one finds $\vec Q_1=-\vec Q_2$ i.e.
$\vec Q_1 \cdot \vec Q_2=-e^2/3$, so that
these flux tubes are attracted each other,
and would be annihilated into dynamical gluons.

\item{\rm (b)}
For the different flux tubes satisfying
$\vec Q_1 \cdot \vec Q_2<0$
(e.g. $R$-$\bar R$ and $B$-$\bar B$),
one finds $\vec Q_1 \cdot \vec Q_2=-e^2/6$,
so that these flux tubes are attractive, and
would be unified into a single flux tube
(similar to $\bar G$-$G$ flux tube).

Based on the above calculation,
we propose a new scenario of the QGP formation via
the annihilation of the color-electric flux
tubes.$^{\ichie}$
When the flux tubes are sufficiently dense
in the central region after the collisions,
many flux tubes are annihilated or unified.
During their annihilation process,
lots of dynamical gluons (and quarks) would be created,
and the energy of the flux tubes turns into that of the
randam kinetic motion of gluons (and quarks).
The thermalization is achieved through the stochastic
gluon collisions, and finally a hot QGP would be created.
Here, the gluon self-interaction in QCD plays an
essential role to the thermalization process,
which is quite different from the photon system in QED.

In more realistic case, both the quark-pair creation and the flux-tube
annihilation would take place at the same time.
For instance, the flux tube breaking $^{\SGTA,\glendenning}$
would occur before the flux tube annihilation for the
dilute flux tube system.
On the other hand,
in case of extremely high energy collisions,
these would be lots of flux tubes overlapping
in the central region between heavy ions,
and therefore the flux tube annihilation
should play the dominant role in the QGP formation.
In any case, the DGL theory would provide a calculable method
for the dynamics of the color-electric flux tubes in the QGP formation.

\chapter{
Relation between Instanton and QCD-monopole Trajectory
}

Finally, we study the relation between the QCD-monopole and
the instanton$^{\rajaraman}$, which is another important
topological object in nonabelian gauge theories.
There is an ambiguity on the gauge-dependent variable $X(x)$
to be diagonalized in the abelian gauge fixing,
and therefore we choose a suitable $X(x)$ to describe the instanton
configuration.
The Polyakov gauge, where $A_4(x)$ is to be diagonalized, is very
interesting, because
$A_4(x)$ takes the hedgehog-like configuration
near the well-localized instanton, and the QCD-monopole
trajectory should pass through its center inevitably.
Here, we show this relation in the Euclidean SU(2)-gauge theory.

The gauge configuration near the well-localized instanton is given by
$$
A_\mu (x) \simeq -i\eta ^{a\mu \nu }\sigma ^a{(x-x_0)^\nu  \over
|(x-x_0)|^2+a_0^2}
\eqn\INST
$$
in the non-singular gauge.$^{\rajaraman}$
Here, $\eta ^{a\mu \nu }$ is the 't~Hooft symbol;
$x_0^\mu  \equiv ({\bf x_0},t_0)$ and $a_0$ denote
the center coordinate and the size of the instanton, respectively.
In particular, one finds
$$
A_4(x) \simeq i{\sigma ^a ({\bf x}-{\bf x_0})^a \over |(x-x_0)|^2+a_0^2},
\eqn\INSTf
$$
so that there inevitably appears a QCD-monopole trajectory
with temporal direction penetrating the center of the instanton
in the Polyakov gauge.
For instance, the QCD-monopole trajectory $x_\mu \equiv({\bf x},t)$
is exactly found as
$$
{\bf x}={\bf x_0}, \hbox{\quad} -\infty <t<\infty
\eqn\MONOt
$$
for the one-instanton solution at the classical level.
Thus, the QCD-monopole trajectory is expected to have a close
relation to the instanton
\REF\miyamuraI{
O.~Miyamura and S.~Origuchi, {\it this Proceedings}.}
configuration.$^{\miyamuraI}$

We find an interesting feature of the QCD-monopole trajectory
in the Polyakov gauge
in the multi-instanton solution$^{\rajaraman}$,
$$
A_\mu (x)=-i\bar \eta ^{a\mu \nu }\sigma ^a
\left(\sum_j {a_j^2 (x-x_j)^\nu  \over |x-x_j|^4}\right)/
\left(1+\sum_k {a_k^2 \over |x-x_k|^2}\right).
\eqn\INSTmany
$$
For instance, there appear two junctions and a loop
in the QCD-monopole trajectory in
the two-instanton system as shown in Fig.4.
Here, we consider the two instantons with the same size
locating at $(\pm x_0,0,0,0)$ for simplicity.
In this case, the QCD-monopole trajectory is found to be
$(x,0,0,t)$ with
$$
x=0  \hbox{\quad or \quad}
t^2=(x_0^2-x^2)+2|x_0|\sqrt{(x_0^2-x^2)}.
\eqn\MONOtra
$$
The QCD-monopole trajectories tend to be highly folded by connecting
their loops in the multi-instanton configuration.
Hence, the presence of instantons is expected to
promote QCD-monopole condensation, which is characterized
by a folded long monopole-loop$^{\kronfeld}$.
This conjecture can be checked by the lattice QCD.

We acknowledge O.~Miyamura and H.~Monden for discussions and comments.
One of the authors (H.S.) is supported by the Special Researchers'
Basic Science Program at RIKEN.

\refout

\figout

\end